\documentclass[aps,pra,preprint,superscriptaddress,nopacs]{revtex4-1}
\usepackage{amsmath,amssymb,latexsym,amsthm,bm}
\usepackage{graphicx}

\begin{document}


\title{Higgs Mode in Superconductors}

\author{Ryo Shimano}
\affiliation{Cryogenic Research Center, The University of Tokyo, Tokyo 113-0032, Japan}
\affiliation{Department of Physics, The University of Tokyo, Tokyo 113-0033, Japan}
\author{Naoto Tsuji}
\affiliation{RIKEN Center for Emergent Matter Science (CEMS), Wako 351-0198, Japan}

\begin{abstract}
When a continuous symmetry of a physical system is spontaneously broken, two types of collective modes typically emerge: 
the amplitude and phase modes of the order-parameter fluctuation. For superconductors, the amplitude mode is recently referred to as the ``Higgs mode'' as it is a condensed-matter analogue of a Higgs boson in particle physics. 
Higgs mode is a scalar excitation of the order parameter, distinct from charge or spin fluctuations, and thus does not couple to electromagnetic fields linearly. This is why the Higgs mode in superconductors has evaded experimental observations over a half century after the initial theoretical prediction, except for a charge-density-wave coexisting system. With the advance of nonlinear and time-resolved terahertz spectroscopy techniques, however, it has become possible to study the Higgs mode through the nonlinear light-Higgs coupling. In this review, we overview recent progresses on the study of the Higgs mode in superconductors.
\end{abstract}


\date{\today}

\maketitle


\section{INTRODUCTION AND A BRIEF HISTORICAL OVERVIEW}

Spontaneous breaking of a continuous symmetry is a fundamental concept of phase transition phenomena 
in various physical systems ranging from condensed matter to high-energy physics \cite{Nambu2011}.
For instance, a ferromagnetic transition is characterized by the spontaneous breaking of spin rotational symmetry. Likewise, superconductivity is characterized by the spontaneous breaking of U(1) rotational symmetry with respect to the phase of a macroscopic wavefunction \cite{Ginzburg1950}.

\begin{figure}
\begin{center}
\includegraphics[width=12.5cm]{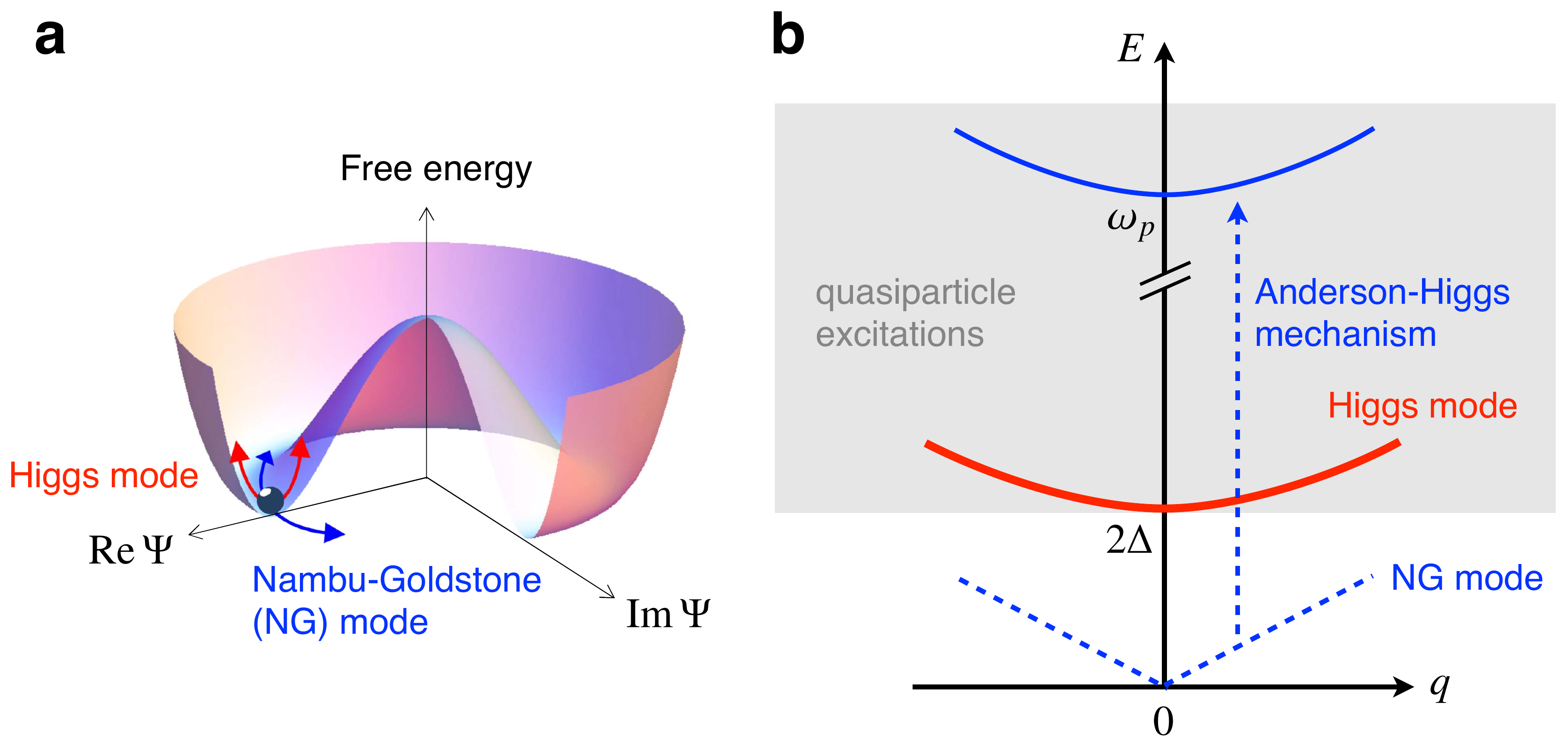}
\caption{({\it a)} A schematic picture of the Higgs (red) and Nambu-Goldstone (blue) modes represented on the 
Mexican-hat free-energy potential as a function of the complex order parameter $\Psi$.
({\it b}) A schematic excitation spectrum of an $s$-wave superconductor.
Due to the Anderson-Higgs mechanism, the Nambu-Goldstone mode acquires an energy gap 
in the order of the plasma frequency $\omega_p$,
while the Higgs mode remains in low energy with an energy gap $2\Delta$,
above which the quasiparticle excitation continuum overlaps.}
\label{Mexican hat}
\end{center}
\end{figure}

When a continuous symmetry is spontaneously broken, there emerge
two types of collective modes in general:
fluctuations of the phase and amplitude of the order parameter as schematically shown in {\bf Figure}~\ref{Mexican hat}\textit{\textbf a}. 
The phase mode (also termed Nambu-Goldstone (NG) mode) is primarily a gapless (massless) mode 
\cite{Bogoliubov1958, Anderson1958a, Anderson1958b, Nambu1960, Goldstone1961, Goldstone1962}
as required by the symmetry.
An acoustic phonon is such an example that is associated 
with spontaneous breaking of translational symmetry in a crystal lattice. 
The amplitude mode, on the other hand, is generally a gapped (massive) mode 
since its excitation costs an energy due to the curvature of the potential ({\bf Figure}~\ref{Mexican hat}\textit{\textbf a}).
The amplitude mode (especially for superconductors) is often called the Higgs mode
according to the close analogy with the Higgs boson in particle physics.
It may sound strange that one talks about the Higgs particle in condensed matter systems,
but in fact the origin of the idea of Higgs physics can be found in the course of the study of superconductivity
\cite{Bogoliubov1958, Anderson1958a, Anderson1958b, Nambu1960, Goldstone1961, Goldstone1962, Anderson1963}.
Namely, the standard model in particle physics can in some sense be viewed as a relativistic version of the Ginzburg-Landau theory \cite{Ginzburg1950},
i.e., a low-energy effective theory of superconductors. The emergence of the Higgs mode is a universal 
and fundamental phenomenon in systems with spontaneous symmetry breaking.

In the case of superconductors where the order parameter couples to gauge fields, 
the Higgs mode occupies a special status since it is the lowest collective excitation mode ({\bf Figure}~\ref{Mexican hat}{\textit{\textbf b}}).
The massless phase mode is absorbed into the longitudinal component of electromagnetic fields,
and is lifted up to high energy in the scale of the plasma frequency due to the Anderson-Higgs mechanism
\cite{Anderson1958a, Anderson1958b, Nambu1960, Anderson1963, Englert1964, Higgs1964a, Higgs1964b, Guralnik1964}. 
As a result, the amplitude mode becomes stable against the decay to the phase mode in superconductors.
The existence of the amplitude mode in superconductors was suggested by Anderson \cite{Anderson1958b, Anderson2015} soon after the development of a microscopic theory of superconductivity by Bardeen, Cooper, and Schrieffer (BCS) \cite{BCS1957}.
In an $s$-wave (BCS-type) superconductor, the Higgs-mode gap energy $\omega_{\rm H}$ coincides 
with the superconducting gap energy $2\Delta$ \cite{Anderson1958b, Volkov1974, Kulik1981, Littlewood1981, Littlewood1982}
(see also \cite{Nambu1961, Nambu1985}).
This is eventually consistent with Nambu's conjectural sum rule \cite{Nambu1985, Volovik2014}: 
$\omega_{\rm H}^2 + \omega_{\rm NG}^2 = 4\Delta^2$ with the NG mode gap energy $\omega_{\rm NG}=0$. 
Since the Higgs mode lies at the lower bound of the quasiparticle continuum near zero momentum, 
the decay of the Higgs mode to single-particle excitations is suppressed and becomes a much slower power law \cite{Volkov1974}.
For nonzero momentum, the Higgs mode can decay by transferring its energy to quasiparticle excitations.
The energy dispersion and the damping rate of the Higgs mode has been derived
within the random phase approximation \cite{Littlewood1981, Littlewood1982}. 


Provided that the theory of the Higgs particle originates from that of superconductivity and that 
the Higgs particle has been discovered in LHC experiments in 2012 \cite{ATLAS2012, CMS2012},
it would be rather surprising that an experimental observation of the Higgs mode in superconductors, a home ground of Higgs physics, 
has been elusive for a long time.
There are, however, good reasons for that: 
The Higgs mode does not have any electric charge, electric dipole, magnetic moment, and other quantum numbers.
In other words, the Higgs mode is a scalar excitation (which is distinct from, e.g., charge fluctuations).
Therefore it does not couple to external probes such as electromagnetic fields in the linear-response regime. 
Another reason is that the energy scale of the Higgs mode lies in that of the superconducting gap, 
which is in the order of millielectron volts in typical metallic superconductors.
To excite the Higgs mode, one needs an intense terahertz (THz) light source (1 THz $\sim$ 4 meV $\sim$ 300 $\mu$m), 
which has become available only in the last decade.

One exception (and the first case) for the observation of the Higgs mode in superconductors in the early stage
was the Raman scattering experiment 
for a superconductor 2H-NbSe$_2$ \cite{Sooryakumar1980, Sooryakumar1981}. It is exceptional in the sense that superconductivity and charge density wave (CDW)
coexist in a single material.
The Raman peak observed near the superconducting gap energy $2\Delta$ in 2H-NbSe$_2$ was first interpreted 
as a single-particle excitation across the gap \cite{Sooryakumar1980, Sooryakumar1981}. 
Soon later, Littlewood and Varma have theoretically elucidated that the peak corresponds to the scalar excitation 
of the amplitude mode (i.e., the Higgs mode) \cite{Littlewood1981, Littlewood1982} (see also Reference~\cite{Varma2002}).
A renewed interest in the Raman signal has recently revealed the importance of the coexisting CDW order for the Higgs mode to be visible 
in the Raman response \cite{Measson2014, Cea2014, Grasset2018}.
Indeed, the Raman peak of the Higgs mode has been shown to be absent in a superconducting NbS$_2$,
which is a material similar to NbSe$_2$ but has no CDW order \cite{Measson2014}. 

It has been a long-standing issue as to whether the Higgs mode can be observed in superconductors without CDW order.
On the theoretical side, there have been various proposals for the excitation of the Higgs mode,
including a quench dynamics \cite{Volkov1974, Barankov2004, Yuzbashyan2005, Barankov2006,
Yuzbashyan2006a, Yuzbashyan2006b, Gurarie2009, Tsuji2013} as well as a laser excitation 
\cite{Papenkort2007, Papenkort2008, Schnyder2011, Krull2014, Tsuji2015, Kemper2015, Chou2017}.
It is only after the development of ultrafast pump-probe spectroscopy techniques in the low-energy terahertz-frequency region
that a clear observation of the Higgs-mode oscillation has been reported in a {\it pure} $s$-wave superconductor NbN \cite{Matsunaga2013}
(with `pure' meaning no other long-range order such as CDW).
Subsequently, it has been experimentally demonstrated 
through the THz pump-probe and third harmonic generation (THG) measurements \cite{Matsunaga2014}
that the Higgs mode can couple to electromagnetic fields in a {\it nonlinear} way.
In the THG experiment, a resonant enhancement of THG was discovered \cite{Matsunaga2014} at the condition $2\omega=2\Delta$ 
with $\omega$ the frequency of the incident THz light, indicating the nonlinear light-Higgs coupling in a two-photon process \cite{Tsuji2015}.

Meanwhile, it has been theoretically pointed out that not only the Higgs mode but also charge density fluctuation (CDF) can contribute 
to the THG signal \cite{Cea2016}. It was shown that the CDF gives a much larger contribution than the Higgs mode 
in a clean-limit superconductor within the BCS mean-field approximation.
While the relative magnitude of the Higgs-mode and CDF contributions to the THG has been under debate
\cite{Cea2016, Tsuji2016, Matsunaga2017, Cea2018}, 
theoretical progresses have recently been made on the theory of the light-Higgs coupling,
showing that both the effects of phonon retardation \cite{Tsuji2016} and nonmagnetic impurity scattering \cite{Jujo2015, Jujo2018, Murotani2019, Silaev2019} drastically enhance the Higgs-mode contribution to nonlinear optical responses, far exceeding the CDF contribution.

The experiments have been extended to high-$T_c$ cuprate superconductors, showing the presence of the Higgs-mode contribution
to the pump-probe signal in $d$-wave superconductors \cite{Katsumi2018}. 
Recently the THG has also been identified in various high-$T_c$ cuprates \cite{Chu2019}.
The Higgs mode can be made visible in a linear optical response if dc supercurrent is flowing in a superconductor \cite{Moor2017}.
The infrared activated Higgs mode in the presence of supercurrents has recently been observed in a superconductor NbN \cite{Nakamura2018}.

In this review article, we overview these recent progresses in the study of the Higgs mode in superconductors 
(where the Anderson-Higgs mechanism is taking place), 
with an emphasis on the experimental aspects. 
In a broader context, collective amplitude modes are not limited to superconductors but are ubiquitous in condensed matter systems
that may not be coupled to gauge fields (hence without the Anderson-Higgs mechanism).
Those include squashing modes in superfluid $^3$He \cite{He3book}, 
amplitude modes in bosonic and fermionic condensates of ultracold-atom systems \cite{Endres2012, Behrle2018}, 
amplitude modes in quantum antiferromagnets \cite{Ruegg2008}, and so on.
A comprehensive review including these topics can be found in Reference~\cite{Pekker2015}. 


\section{NONLINEAR LIGHT-HIGGS COUPLING}
\label{nonlinear coupling}

In this section, we review from a theoretical point of view how the Higgs mode in superconductors 
couples to electromagnetic fields in a {\it nonlinear} way. It is important to understand the mechanism 
of the nonlinear light-Higgs coupling for observing the Higgs mode in experiments. 
In fact, the Higgs mode does not have a linear response against electromagnetic fields in usual situations,
which has been an obstacle for experiments to detect the Higgs mode in superconductors for a long time.

\subsection{A phenomenological view}
\label{phenomenological view}

Let us first take a phenomenological point of view based on
the Ginzburg-Landau (GL) theory \cite{Ginzburg1950}, which provides us with a quick look at the nonlinear
light-Higgs coupling. For early developments based on the time-dependent GL theory, 
we refer to References~\cite{Abraham1966, Schmid1966, Caroli1967, Ebisawa1971, SadeMelo1993}.
In the GL theory, the free energy density $f$ is assumed to be
a function of a complex order-parameter field $\psi({\bm r})$,
\begin{eqnarray}
f[\psi]
&=
\displaystyle
f_0+a|\psi({\bm r})|^2+\frac{b}{2}|\psi({\bm r})|^4
+\frac{1}{2m^\ast}|(-i\nabla-e^\ast{\bm A})\psi({\bm r})|^2,
\label{GL}
\end{eqnarray}
where $a=a_0(T-T_c)$, $a_0$ and $b$ are some constants,
$m^\ast$ and $e^\ast$ are the effective mass and the effective charge of the Cooper pair condensate,
and ${\bm A}$ is the vector potential that represents the external light field ($\bm E(t)=-\partial \bm A(t)/\partial t$). 
The amplitude of the order parameter corresponds to the superfluid density (i.e., $|\psi|^2=n_s$),
while the phase corresponds to that of the condensate.

The free energy density (\ref{GL})
is invariant under 
the global U(1) phase rotation $\psi({\bm r})\to e^{i\varphi}\psi({\bm r})$
and more generally under the gauge transformation $\psi({\bm r})\to e^{ie^\ast \chi({\bm r})}\psi({\bm r})$,
${\bm A}({\bm r})\to {\bm A}({\bm r})+\nabla \chi({\bm r})$ for an arbitrary scalar field $\chi({\bm r})$.
In addition, the GL free energy density (\ref{GL}) is invariant against
the particle-hole transformation $\psi({\bm r})\to \psi^\dagger({\bm r})$. The presence of the particle-hole symmetry
(including the time derivative terms in the action)
is crucial \cite{Varma2002, Pekker2015, Tsuchiya2018} in decoupling
the Higgs amplitude mode from the phase mode (Nambu-Goldstone mode) 
in electrically neutral systems such as superfluid ultracold atoms.
Without the particle-hole symmetry, the Higgs mode quickly decays into the phase mode and becomes short-lived.
In superconductors (which consist of electrically charged electrons), on the other hand, the phase mode acquires an energy gap
in the order of the plasma frequency due to the Anderson-Higgs mechanism.
This prevents the Higgs mode
from decaying into the phase mode. The particle-hole symmetry itself is realized as an approximate symmetry
in the Bogoliubov-de Gennes Hamiltonian, which describes microscopic low-energy physics of superconductors
around the Fermi energy at the mean-field level.

When the temperature goes below the critical temperature (i.e., $a<0$), the global U(1) symmetry is spontaneously broken,
and the system turns into a superconducting state. The order-parameter fluctuation from the ground state ($\psi({\bm r})=\psi_0$)
can be decomposed
into amplitude $H({\bm r})$ and phase $\theta({\bm r})$ components,
\begin{eqnarray}
\psi({\bm r})
&=
[\psi_0+H({\bm r})]e^{i\theta({\bm r})}.
\end{eqnarray}
Then, the GL potential can be expressed up to the second order of the fluctuation as
\begin{eqnarray}
f
&=
\displaystyle
-2aH^2+\frac{1}{2m^\ast}(\nabla H)^2
+\frac{e^{\ast 2}}{2m^\ast}
\left({\bm A}-\frac{1}{e^\ast}\nabla\theta\right)^2(\psi_0+H)^2+\cdots.
\label{GL broken}
\end{eqnarray}
The first term on the right hand side of Equation~\ref{GL broken} indicates
that the Higgs mode has an energy gap ({\bf Figure}~\ref{Mexican hat}{\textit{\textbf b}) proportional to $(-a)^{1/2}\propto (T_c-T)^{1/2}$. 
The microscopic BCS mean-field theory predicts that the Higgs gap is actually identical to the superconducting gap $2\Delta$
\cite{Anderson1958b, Volkov1974, Kulik1981, Littlewood1981, Littlewood1982}.
The phase field $\theta({\bm r})$
does not have such a mass term, so that at first sight one would expect to have a massless Nambu-Goldstone mode.
However, $\theta({\bm r})$ always appears in the form of $({\bm A}-\nabla \theta/e^\ast)$ in Equation~\ref{GL broken}, which
allows one to eliminate $\theta({\bm r})$ from the GL potential by taking a unitary gauge ${\bm A}'={\bm A}-\nabla\theta/e^\ast$.
As a result, we obtain (denoting ${\bm A}'$ as ${\bm A}$)
\begin{eqnarray}
f
&=
\displaystyle
-2aH^2+\frac{1}{2m^\ast}(\nabla H)^2
+\frac{e^{\ast 2}\psi_0^2}{2m^\ast}{\bm A}^2
+\frac{e^{\ast 2}\psi_0}{m^\ast}{\bm A}^2 H
+\cdots.
\label{GL broken2}
\end{eqnarray}
One can see that the phase mode is absorbed into the longitudinal component of the electromagnetic field.
At the same time, there appears a mass term for photons (the third term on the right hand side of Equation \ref{GL broken2})
with the mass proportional to $\sqrt{e^{\ast 2}\psi_0^2/m^\ast}$ (Anderson-Higgs mechanism). Due to this, electromagnetic waves cannot 
propagate freely inside superconductors
but decay exponentially with a finite penetration length (Meissner effect).

From Equation~\ref{GL broken2}, one immediately sees that there is no linear coupling between the Higgs and electromagnetic field.
This is consistent with the fact that the Higgs mode does not have an electric charge, magnetic moment, and other quantum numbers, which has been a main obstacle in observing the Higgs mode by an external probe for a long time.
On the other hand, there is a nonlinear coupling term ${\bm A}^2 H$ in Equation~\ref{GL broken2}, which is responsible
for the Higgs mode to contribute in various nonlinear processes. In {\bf Figure}~\ref{higgs-diagram}{\textit{\textbf a}}, we illustrate the diagrammatic representation
of the nonlinear light-Higgs coupling, where two photons are coming in with frequencies $\omega_1$ and $\omega_2$
and one Higgs is emitted with frequency $\omega_1+\omega_2$.
This is analogous to the elementary process of the Higgs particle decaying into W bosons that had played a role
in the discovery of the Higgs particle in LHC experiments \cite{ATLAS2012, CMS2012}.

\begin{figure}
\includegraphics[width=12.5cm]{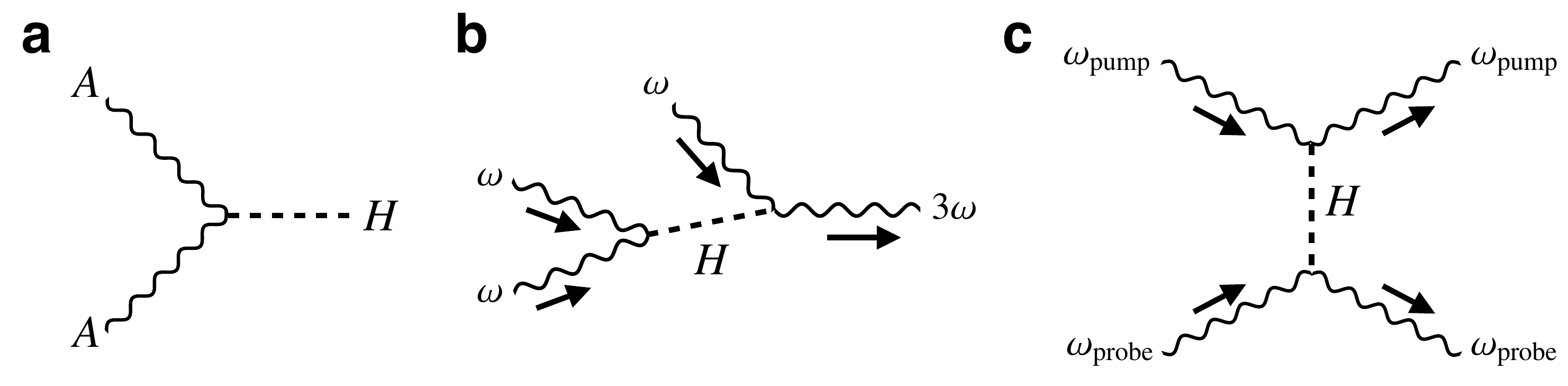}
\caption{({\it a}) Diagrammatic representation of the nonlinear light-Higgs coupling $\bm A^2 H$ in Equation~\ref{GL broken2}.
({\it b}) Third harmonic generation mediated by the Higgs mode. ({\it c}) Pump-probe spectroscopy mediated by the Higgs mode.}
\label{higgs-diagram}
\end{figure}

Due to the nonlinear interaction between the Higgs mode and electromagnetic fields,
one can induce a nonlinear current given by
\begin{eqnarray}
{\bm j}
&=
\displaystyle
-\frac{\partial F}{\partial \bm A}
=
-\frac{ie^\ast}{2m^\ast}[\psi^\dagger \nabla \psi-(\nabla\psi^\dagger) \psi]
-\frac{e^{\ast 2}}{m^\ast}{\bm A}\psi^\dagger \psi.
\label{current}
\end{eqnarray}
Again, we expand $\psi$ around the ground state $\psi_0$ in Equation~\ref{current} and collect leading terms in $H$, which results in
\begin{eqnarray}
{\bm j}
&=
\displaystyle
-\frac{2e^{\ast 2}\psi_0}{m^\ast}{\bm A}H.
\label{London}
\end{eqnarray}
This is nothing but the leading part of the London equation $\bm j=-(e^{\ast 2}n_s/m^\ast)\bm A$
(let us recall that the amplitude of the order parameter corresponds to the superfluid density, $|\psi|^2=n_s$).

When a monochromatic laser with frequency $\omega$ drives a superconductor, the amplitude of the order parameter
oscillates with frequency $2\omega$ due the nonlinear coupling $\bm A^2 H$ ({\bf Figure}~\ref{higgs-diagram}{\textit{\textbf a}}).
Together with the oscillation of $\bm A$ with frequency $\omega$ in Equation~\ref{London}, the induced nonlinear current
shows an oscillation with frequency $3\omega$. As a result, one obtains the third harmonic generation
mediated by the Higgs mode ({\bf Figure}~\ref{higgs-diagram}{\textit{\textbf b}}). Since the Higgs mode has an energy $2\Delta$ at low momentum, one can resonantly
excite the Higgs mode by a THz laser with a resonance condition $2\omega=2\Delta$. 
Given that the nonlinear current is proportional to the Higgs-mode amplitude (Equation~\ref{London}),
the third harmonic generation can be resonantly enhanced through the excitation of the Higgs mode.
The Higgs-mode resonance in third harmonic generation
has been experimentally observed for a conventional superconductor \cite{Matsunaga2014}.

Another example of nonlinear processes to which the Higgs mode can contribute is a pump-probe spectroscopy
where pump and probe light is simultaneously applied
({\bf Figure}~\ref{higgs-diagram}{\textit{\textbf c}}). In this case, photons with different frequencies $\omega_{\rm pump}$ and $\omega_{\rm probe}$
are injected, and those with the same respective frequencies are emitted. There are three possibilities of the frequency carried by
the Higgs mode: $\omega_{\rm pump}\pm \omega_{\rm probe}$ and 
$\omega_{\rm pump}-\omega_{\rm pump}=\omega_{\rm probe}-\omega_{\rm probe}=0$.
If one uses a THz pump and optical probe, i.e., $\omega_{\rm pump}\lesssim 2\Delta$ and $\omega_{\rm probe}\gg 2\Delta$, 
the first two channels do not contribute since the excitation
of the Higgs mode is far off-resonant. The remaining zero-frequency excitation channel contributes to the pump-probe spectroscopy.
This process has been elaborated in the experimental study of the Higgs mode in $d$-wave superconductors \cite{Katsumi2018}.

\subsection{A microscopic view}
\label{microscopic view}

In the previous subsection, we reviewed the phenomenology for the nonlinear light-Higgs coupling using
the Ginzburg-Landau theory. However, strictly speaking, 
the application of the Ginzburg-Landau theory to nonequilibrium problems
is not microscopically justified in the case of gapped superconductors \cite{Gorkov1968, Gulian1999, Kopnin2001}.
The reason is that in a usual situation 
one cannot neglect the effect of quasiparticle excitations whose relaxation time
is longer than the time scale of the order-parameter variation.
Moreover, the Higgs mode and quasiparticle excitations are energetically degenerate at low momentum
({\bf Figure}~\ref{Mexican hat}{\textit{\textbf b}}), so that it is inevitable to excite quasiparticles at the same time
when one excites the Higgs mode. 
This motivates us to take a microscopic approach for further understanding.

\begin{figure}
\begin{center}
\includegraphics[width=12.5cm]{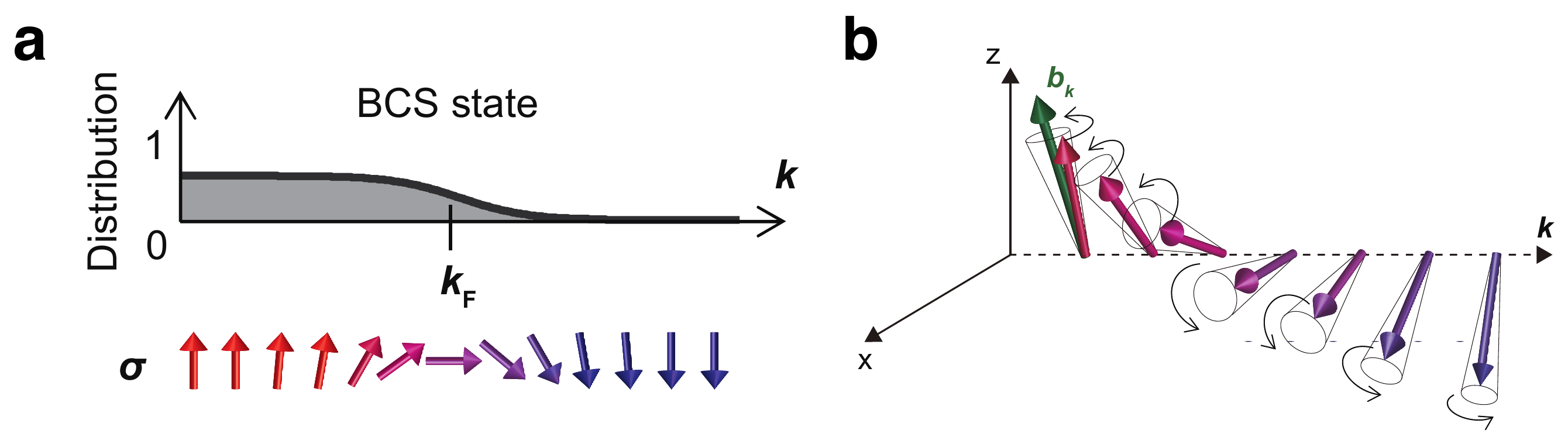}
\caption{({\it a}) Momentum distribution of a superconducting state (above)
and its pseudospin representation due to Anderson (below).
({\it b}) Pseudospin precession induced by a laser field. Figures are taken from Reference~\cite{Matsunaga2014}.}
\label{pseudospin}
\end{center}
\end{figure}

Let us adopt the time-dependent BCS or Bogoliubov-de Gennes equation,
which is efficiently represented by the Anderson pseudospin
$\bm \sigma_{\bm k}
=
\frac{1}{2}\langle\Psi_{\bm k}^\dagger \bm\tau \Psi_{\bm k}\rangle$ \cite{Anderson1958b}.
Here $\Psi_{\bm k}^\dagger=(c_{\bm k\uparrow}^\dagger, c_{-\bm k\downarrow})$ is the Nambu spinor, 
$\bm \tau=(\tau_x,\tau_y,\tau_z)$ are Pauli matrices, $c_{\bm k\sigma}^\dagger$ ($c_{\bm k\sigma}$)
is the creation (annihilation) operator of electrons with momentum $\bm k$ and spin $\sigma$,
and $\langle \cdots \rangle$ denotes the statistical average.
Physically, the $x$ and $y$ components of the Anderson pseudospin correspond to the real and imaginary parts
of the Cooper pair density, respectively, and the $z$ component corresponds to the 
momentum occupation distribution of electrons ({\bf Figure}~\ref{pseudospin}{\textit{\textbf a}}). With this notation, the BCS mean-field Hamiltonian is written as
\begin{align}
H_{\rm BCS}
&=
2\sum_{\bm k}\bm b_{\bm k}(t)\cdot\bm \sigma_{\bm k},
\\
\bm b_{\bm k}(t)
&=
\left(-\Delta'(t), -\Delta''(t), \frac{1}{2}(\epsilon_{\bm k+\bm A(t)}+\epsilon_{\bm k-\bm A(t)})\right).
\end{align}
Here $\bm b_{\bm k}(t)$ is a pseudomagnetic field acting on pseudospins, $\Delta'$ and $\Delta''$ are
the real and imaginary parts of the superconducting gap function, respectively, 
and $\epsilon_{\bm k}$ is the band dispersion of the system.
The equation of motion for the pseudospins is given by the Bloch-type equation,
\begin{align}
\frac{\partial}{\partial t}\bm\sigma_{\bm k}(t)
&=
2\bm b_{\bm k}(t)\times\bm \sigma_{\bm k}(t),
\label{Bloch}
\end{align}
supplemented by the mean-field condition $\Delta'(t)+i\Delta''(t)=\frac{V}{N}\sum_{\bm k}[\sigma_{\bm k}^x(t)+i\sigma_{\bm k}^y(t)]$
with $V$ the attractive interaction and $N$ the number of $k$ points.
The time evolution of the superconducting state is thus translated into
the precession dynamics of Anderson pseudospins ({\bf Figure}~\ref{pseudospin}{\textit{\textbf b}}).

\begin{figure}
\begin{center}
\includegraphics[width=12cm]{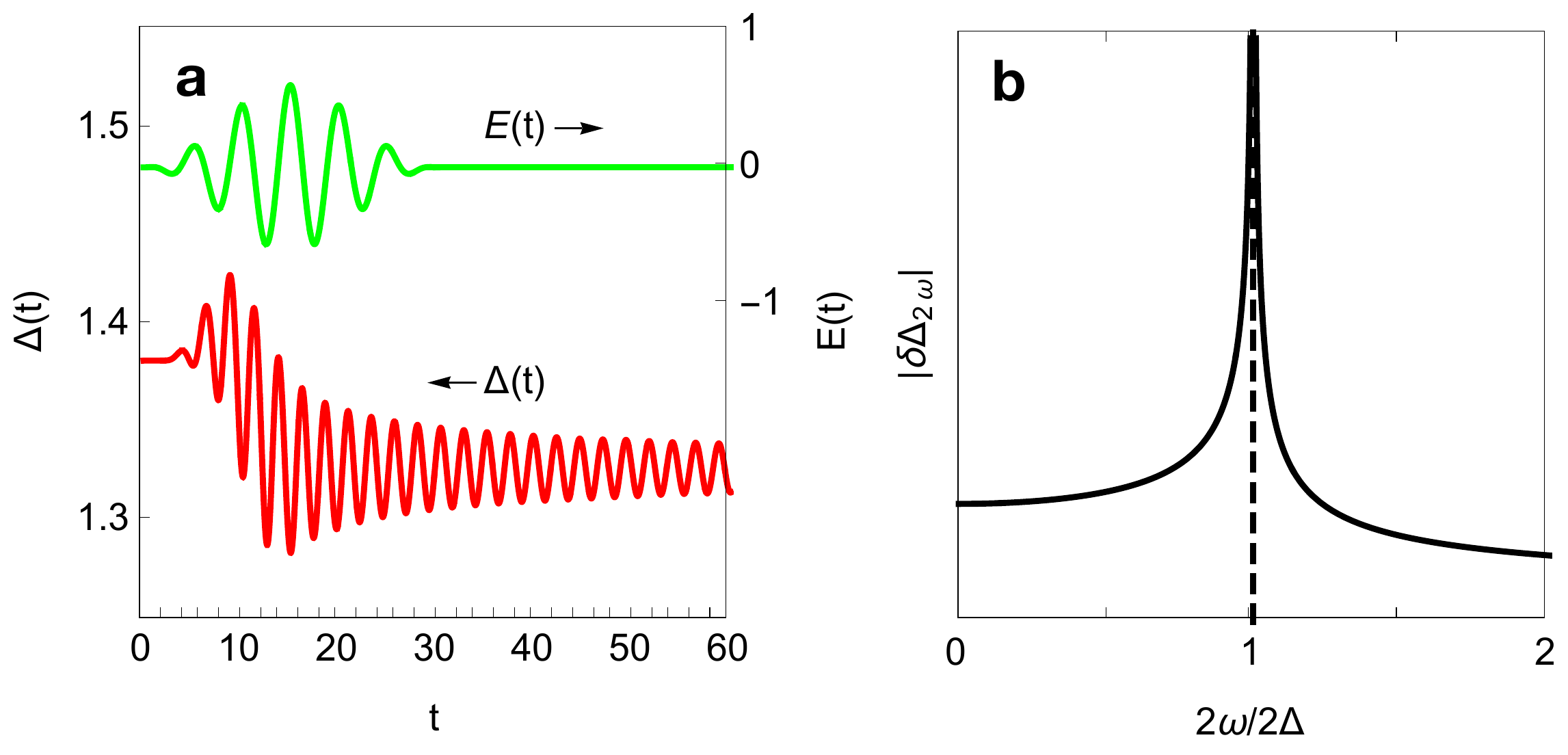}
\caption{({\it a}) A numerical result for the time evolution of the superconducting gap $\Delta(t)$
driven by an electric field pulse $E(t)$ with the frequency $\omega=2\pi/5\sim 1.26$.
Here we take a two dimensional square lattice with a bandwidth 8 at half filling. The interaction is $V=4$, and 
the initial temperature is $T=0.02$. The polarization of the electric field is parallel to $x$ axis.
The left (right) vertical axis represents the value of $\Delta(t)$ ($E(t)$) as indicated by the arrows.
({\it b}) The amplitude of the $2\omega$ oscillation of the superconducting gap as a function of $2\omega/2\Delta$.
({\it b}) is reproduced from Reference~\cite{Tsuji2015}.}
\label{gap}
\end{center}
\end{figure}

The coupling to the electromagnetic field appears in the $z$ component of the pseudomagnetic field,
$b_{\bm k}^z=\frac{1}{2}(\epsilon_{\bm k+\bm A(t)}+\epsilon_{\bm k-\bm A(t)})$.
If we expand $b_{\bm k}^z$ in terms of $\bm A(t)$, we obtain
$b_{\bm k}^z=\epsilon_{\bm k}+\frac{1}{2}\sum_{ij} \frac{\partial^2 \epsilon_{\bm k}}{\partial k_i\partial k_j}A_iA_j+O(A^4)$.
The linear coupling term vanishes, which is consistent with the phenomenology that we have reviewed
in the previous subsection.
The leading nonlinear coupling term is quadratic in $\bm A(t)$. 
In {\bf Figure}~\ref{gap}{\textit{\textbf a}}, we show a numerical result for the time evolution of the gap function $\Delta(t)$
when the system is driven by a multi-cycle electric-field pulse.
One can see that the $2\omega$ oscillation of the gap is generated during the pulse irradiation,
after which a free gap oscillation continues with the frequency $2\Delta$ and the amplitude slowly damping as $t^{-1/2}$
\cite{Volkov1974}.

For a monochromatic wave $\bm A(t)=\bm A_0 \sin \omega t$,
one can solve the equation of motion semi-analytically
by linearizing with respect to the quadratic coupling $A_i(t)A_j(t)$ around the equilibrium solution.
The result shows that the amplitude of the gap function varies as \cite{Tsuji2015}
\begin{align}
\delta\Delta(t)
&\propto
|2\omega-2\Delta|^{-\frac{1}{2}} \cos(2\omega-\phi)
\label{delta Delta}
\end{align}
for $\omega \sim \Delta$,
where $\phi$ is a phase shift. The induced oscillation frequency is $2\omega$, reflecting the nonlinear coupling between the Higgs mode and electromagnetic fields,
$\bm A^2 H$. The oscillation amplitude diverges when the condition $2\omega=2\Delta$ is fulfilled ({\bf Figure}~\ref{gap}{\textit{\textbf b}}). 
This is much the same as the spin resonance phenomenon:
The forced collective precession of the pseudospins with frequency $2\omega$ due to the nonlinear coupling resonates with
the Higgs mode whose energy coincides with $2\Delta$.
The power of the divergence $\frac{1}{2}$ in Equation~\ref{delta Delta} is smaller than that of a resonance with an infinitely long-lived mode
($\sim |2\omega-\omega_\ast|^{-1}$). The reason is that the precession gradually dephases
due to the momentum-dependent frequency $\omega_{\bm k}=2\sqrt{\epsilon_{\bm k}^2+\Delta^2}$ for each pseudospin
so that the average of the pseudospins $\delta\Delta(t)=\frac{V}{N}\sum_{\bm k} \delta\sigma_{\bm k}^x(t)$
shows a power-law damping of oscillations. The phase shift $\phi$ exhibits a $\frac{\pi}{2}$ jump 
as one goes across the resonance by changing the drive frequency $\omega$.

In the case of the square lattice and the polarization of the electric field parallel to the diagonal direction in the $xy$ plane,
for example, 
the current is given by $\bm j(t)\propto \bm A(t)\delta\Delta(t)$ \cite{Tsuji2015, Matsunaga2014},
which corresponds to the phenomenological relation $\bm j(t)\propto \bm A(t)H(t)$
(Equation~\ref{London}). As expected, the Higgs mode induces a resonant enhancement of the third harmonic generation.
Later, it has been pointed out \cite{Cea2016} that the current relation is valid only in rather special situations such as
the square lattice and the polarization parallel to the diagonal direction. In general, the BCS mean-field treatment (in the clean limit) 
suggests that the Higgs-mode contribution
to the third harmonic generation is subdominant as compared to that of individual quasiparticle excitations.
Since the energy scales of the Higgs mode and the quasiparticle pair excitations are in the same order ($\sim 2\Delta$, see {\bf Figure}~\ref{Mexican hat}{\textit {\textbf b}}), the competition between the two contributions always matters.
This point will be further examined in the next subsection.

\subsection{Impurity and phonon assisting}
\label{impurity}

As we have seen in the previous subsection, the Higgs-mode contribution to the third harmonic generation is generally subdominant
in the BCS clean limit.
However, there is a growing understanding from recent studies
\cite{Tsuji2016, Matsunaga2017, Jujo2015, Jujo2018, Murotani2019, Silaev2019}
that if one takes into account effects beyond the BCS mean-field theory in the clean limit
(such as impurity scattering and phonon retardation) 
the light-Higgs coupling strength drastically changes.

To understand the impact of these effects, let us consider the linear optical response.
The current induced by a laser field is decomposed into the paramagnetic and diamagnetic components
$\bm j=\bm j_{\rm para}+\bm j_{\rm dia}$ with \cite{SchriefferBook}
\begin{align}
\bm j_{\rm para}
&=
\sum_{\bm k\sigma} \frac{\partial \epsilon_{\bm k}}{\partial \bm k}\langle c_{\bm k\sigma}^\dagger c_{\bm k\sigma}\rangle,
\\
\bm j_{\rm dia}
&=
-\sum_{\bm k\sigma i} \frac{\partial^2 \epsilon_{\bm k}}{\partial \bm k\partial k_i}A_i\langle c_{\bm k\sigma}^\dagger c_{\bm k\sigma}\rangle.
\end{align}
As we have seen in the pseudospin picture in the previous subsection, only the diamagnetic coupling
contributes to the optical response in the mean-field level with no disorder, 
resulting in ${\rm Re}\,\sigma(\omega)\propto \delta(\omega)$ and ${\rm Im}\,\sigma(\omega)\propto 1/\omega$
for the superconducting state.
For finite frequency ($\omega\neq 0$), the real part of the optical conductivity vanishes in the BCS clean limit.
This does not necessarily mean that the real part of the optical conductivity is suppressed in real experimental situations.
For example, in superconductors with a disorder the real part of the optical conductivity is nonzero
and not even small for $\omega\neq 0$.
In fact, a superconductor NbN used in the experiment of the third harmonic generation turns out to be close to the dirty regime
($\gamma\gg 2\Delta$, where $\gamma$ is the impurity scattering rate) as confirmed
from the measurement of the optical conductivity \cite{Matsunaga2013}.

\begin{table}[h]
\tabcolsep7.5pt
\caption{Relative order of magnitudes of the third-order current $\bm j^{(3)}$ in general situations \cite{Murotani2019}.}
\label{j3}
\begin{center}
\begin{tabular}{c|c|c}
\hline
mode & channel & clean $\to$ dirty \\
\hline
\hline
Higgs & dia ($\bm A^2$) & $(\Delta/\epsilon_F)^2$ \\
 & para ($\bm p\cdot\bm A$) & $(\epsilon_F\gamma/\Delta^2)^2 \to (\epsilon_F/\gamma)^2$ \\
\hline
Quasiparticles & dia ($\bm A^2$) & 1 \\
 & para ($\bm p\cdot\bm A$) & $(\epsilon_F\gamma/\Delta^2)^2 \to (\epsilon_F/\gamma)^2$ \\
\hline
\end{tabular}
\end{center}
\end{table}

In much the same way as the optical conductivity, the nonlinear response intensity is strongly affected by the impurity effect.
The $\gamma$ dependence of the relative order of magnitudes of the third-order current is summarized in Table \ref{j3}, 
where $\epsilon_F$ is the Fermi energy
and the unit of $\bm j^{(3)}$ is taken such that the diamagnetic-coupling contribution of the quasiparticles is set to be of order 1
(which does not significantly depend on $\gamma$).
In the BCS clean limit ($\gamma\to 0$), we have only the diamagnetic coupling, for which the Higgs-mode contribution is 
subleading in the order of $(\Delta/\epsilon_F)^2$ compared to the quasiparticles. 
However, in the presence of impurities ($\gamma\neq 0$)
the paramagnetic coupling generally emerges to contribute to the third harmonic generation.
Its relative order of magnitude changes as $(\epsilon_F\gamma/\Delta^2)^2 \to (\epsilon_F/\gamma)^2$
from the clean to dirty limit. The maximum strength of the paramagnetic coupling is achieved around $\gamma\sim\Delta$,
where the relative order reaches $(\epsilon_F/\Delta)^2$ for both the Higgs mode and quasiparticles.
Thus, the impurity scattering drastically enhances
the light-Higgs coupling: in the clean limit the Higgs-mode contribution is subleading to quasiparticles, whereas in the dirty regime
it becomes comparable to or even larger than the quasiparticle contribution.
The precise ratio between the Higgs and quasiparticle contributions may depend on details of the system.
Recent studies \cite{Jujo2018, Murotani2019, Silaev2019}
suggest that in the dirty regime the Higgs-mode contribution
to THG is an order of magnitude larger than the quasiparticle contribution.
An enhancement of the paramagnetic coupling also occurs for strongly coupled superconductors due to 
the phonon retardation effect \cite{Tsuji2016}.

\subsection{Further developments}

So far, we have reviewed the phenomenological and mean-field treatments
of the nonlinear light-Higgs coupling for conventional $s$-wave superconductors with or without disorder.
For strongly coupled superconductors with an electron-phonon coupling constant $\lambda\gtrsim 1$
(which is the case for NbN superconductors \cite{Kihlstrom1985, Brorson1990, Chockalingam2008}), 
one has to take account of strong correlation effects. One useful approach for this
is the nonequilibrium dynamical mean-field theory \cite{Aoki2014},
which takes into account local dynamical correlations by mapping a lattice model into a local impurity model
embedded in an effective mean field. With this, the Higgs mode in strongly coupled superconductors
modeled by the Holstein model has been analyzed \cite{Tsuji2015, Tsuji2016, Murakami2016a, Murakami2016b}.
A closely related nonequilibrium Keldysh method has been employed to study the Higgs mode in
a time-resolved photoemission spectroscopy \cite{Kemper2015, Nosarzewski2017} and time-resolved optical conductivity \cite{Kumar2019}.
Another approach is to use the gauge-invariant kinetic equation \cite{Yu2017a, Yu2017b, Yang2018, Yang2018b},
where the shift of the center-of-mass momentum of Cooper pairs (``drive effect'') caused by
a laser pulse has been emphasized.

For multi-band superconductors with multiple gaps such as MgB$_2$,
there are not only Higgs modes corresponding to amplitude oscillations of multiple order parameters
but also the so-called Leggett mode \cite{Leggett1966}, 
which corresponds to a collective oscillation of the relative phase between different order parameters.
Based on the mean-field theory, possible laser excitations of the Higgs and Leggett modes in multiband superconductors 
have been studied
\cite{Murotani2019, Akbari2013, Krull2016, Cea2016b, Murotani2017}.

For non-$s$-wave superconductors, one can expect even richer dynamics of the order parameters.
For example, $d$-wave superconductors allow for various different symmetries of amplitude modes \cite{Barlas2013}.
The quench dynamics of the $d$-wave \cite{Peronaci2015} and $p+ip$-wave \cite{Foster2013} Higgs modes 
have been explored.
For recent experimental progresses on pump-probe spectroscopies and third harmonic generation in $d$-wave superconductors, 
we refer to Section~\ref{cuprate}.
It has also been proposed \cite{Fauseweh2017} 
that by measuring Higgs modes for various quench symmetries one can obtain
information on the symmetry of the gap function (``Higgs spectroscopy'').

\section{EXPERIMENTS}

In this section, we overview recent progresses on experimental observations of the Higgs mode in superconductors.
Those include Raman experiments in CDW-coexisting superconductors 2H-NbSe$_2$ and TaS$_2$ (Section \ref{CDW}),
THz spectroscopies and third harmonic generation in a pure $s$-wave superconductor NbN (Section \ref{pure s-wave})
and in high-$T_c$ cuprates (Section \ref{cuprate}), and THz transmittance experiments in NbN
under supercurrent injection (Section \ref{supercurrent}).

\subsection{Higgs mode in a superconductor with CDW}
\label{CDW}

The pioneering work on the observation of the Higgs mode in superconductors dates back to 1980 when a Raman experiment was performed in 2H-NbSe$_2$, a transition metal dichalcogenide in which superconductivity coexists with CDW \cite{Sooryakumar1980, Sooryakumar1981}. A new peak was observed below $T_c$, distinct from the amplitude mode associated with the CDW order. 
Although it was initially recognized as a pair breaking peak, soon after it was identified as the collective amplitude mode of the superconducting order (i.e., the Higgs mode) \cite{Littlewood1981, Littlewood1982}
(see also \cite{Varma2002, Pekker2015}). 

A renewed Raman experiment has been conducted recently, revealing the transfer of the oscillator strength from 
the amplitude mode of CDW (amplitudon) to the Higgs mode \cite{Measson2014}.
Subsequently the Raman experiment under hydrostatic pressure has been performed in 2H-NbSe$_2$ \cite{Grasset2018} ({\bf Figure}~\ref{Raman}). With applying the hydrostatic pressure the CDW order is suppressed and concomitantly the Raman peak identified as the Higgs mode was shown to disappear with leaving only the Cooper pair breaking peak. A more clear separation between the in-gap Higgs mode and the pair breaking peak was recently reported in a similar CDW-coexisting superconductor, TaS$_2$ \cite{Grasset2018b}. These results indicate that the existence of CDW plays an important role on the visibility of the Higgs mode in the Raman spectrum. A theoretical study has demonstrated that, due to the coupling of the superconducting order to the coexisting CDW order, the Higgs mode energy is pushed down below the superconducting gap $2\Delta$ \cite{Cea2014}. Accordingly, the Higgs mode becomes stable due to the disappearance of the decay channel into the quasiparticle continuum. 

\begin{figure}
\begin{center}
\includegraphics[width=12cm]{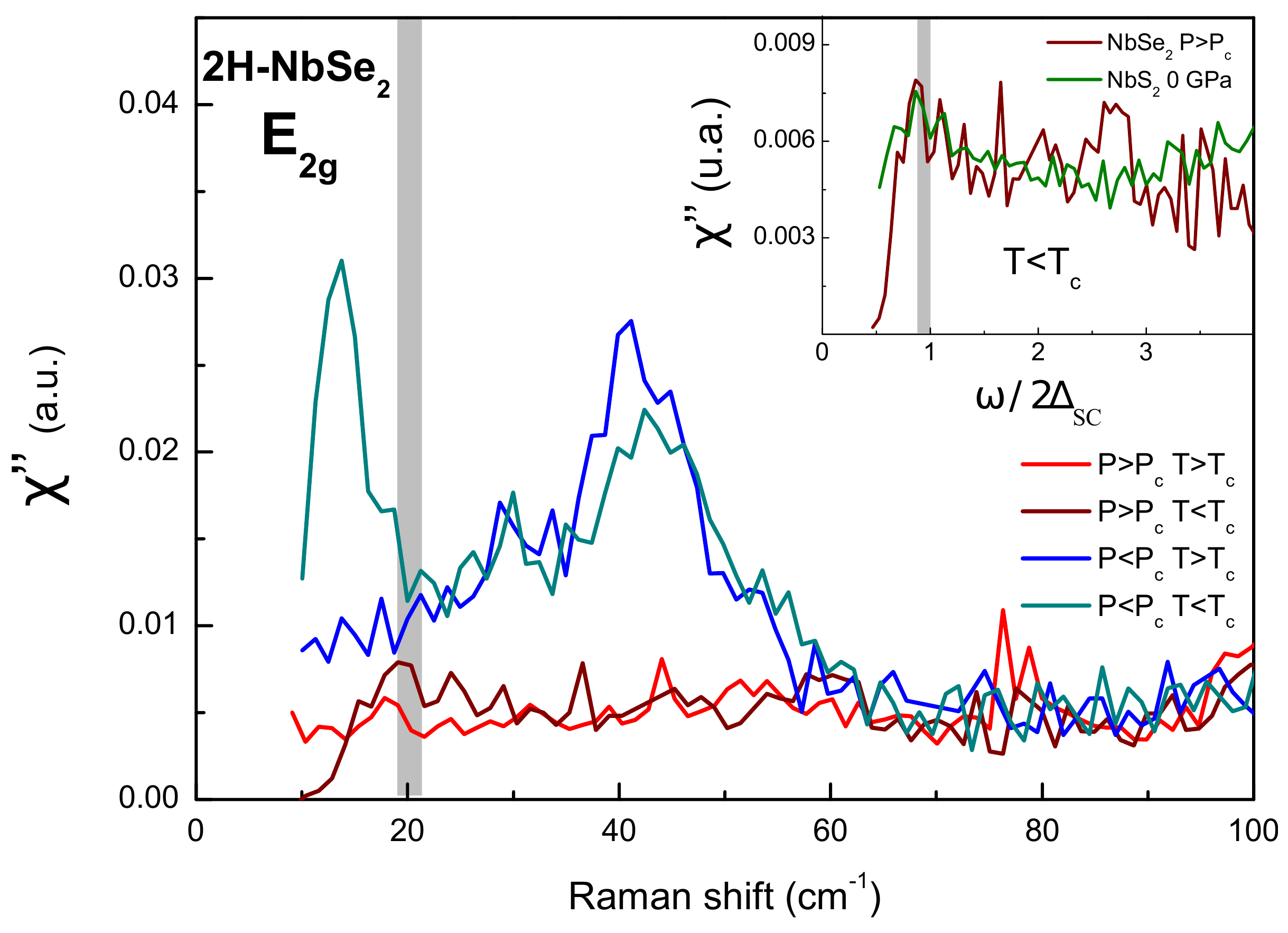}
\caption{Raman spectra in E$_{\rm 2g}$ symmetry of 2H-NbSe$_2$ measured at various temperatures and pressures. Under the ambient pressure ($P<P_c$) and in the superconducting phase ($T<T_c$) (green), a sharp peak identified as the Higgs mode appears below the superconducting gap 2$\Delta\sim 20$ cm$^{-1}$ marked by the gray vertical line. In the CDW phase ($T>T_c$) (blue), only the CDW amplitude mode is observed around 40 cm$^{-1}$. Under the pressure ($P>P_c$) where CDW collapses, only a pair breaking peak is discerned  at $T<T_c$ (brown), and no peak is identified at $T>T_c$ (red). Inset: Raman spectra in the superconducting state without CDW in 2H-NbSe$_2$ (above 4 GPa) and non-CDW-coexisting NbS$_2$ (0 GPa), both of which show only the pair breaking peak.
Frequency is normalized by $2\Delta$. Figure is taken from Reference \cite{Grasset2018}.}
\label{Raman}
\end{center}
\end{figure}

\subsection{Higgs mode in a pure $s$-wave superconductor}
\label{pure s-wave}

\subsubsection{Nonadiabatic quench with a single-cycle THz pump pulse}

One way to excite the Higgs mode is a nonadiabatic quench of the superconducting state,
which can be induced by, for example, a sudden change of the pairing interaction
that generates the order-parameter dynamics,
\begin{eqnarray}
\frac{|\Delta(t)|}{\Delta_{\infty}}
&\simeq
\displaystyle
1+a\frac{\cos(2\Delta_{\infty}t+\phi)}{\sqrt{\Delta_{\infty} t}},
\label{quench dynamics}
\end{eqnarray}
at long time. Here $2\Delta_{\infty}$ is the long-time asymptotic superconducting gap, $a$ is some constant, and $\phi$ is a phase shift.
The behavior (Equation \ref{quench dynamics}) has been theoretically derived
on the basis of the time-dependent BCS mean-field theory \cite{Volkov1974, Barankov2006,
Yuzbashyan2006a, Yuzbashyan2006b}. A similar dynamics has also been investigated after a laser-pulse excitation 
\cite{Papenkort2007, Papenkort2008, Schnyder2011, Krull2014, Kemper2015, Chou2017}.

\begin{figure}
\begin{center}
\includegraphics[width=12cm]{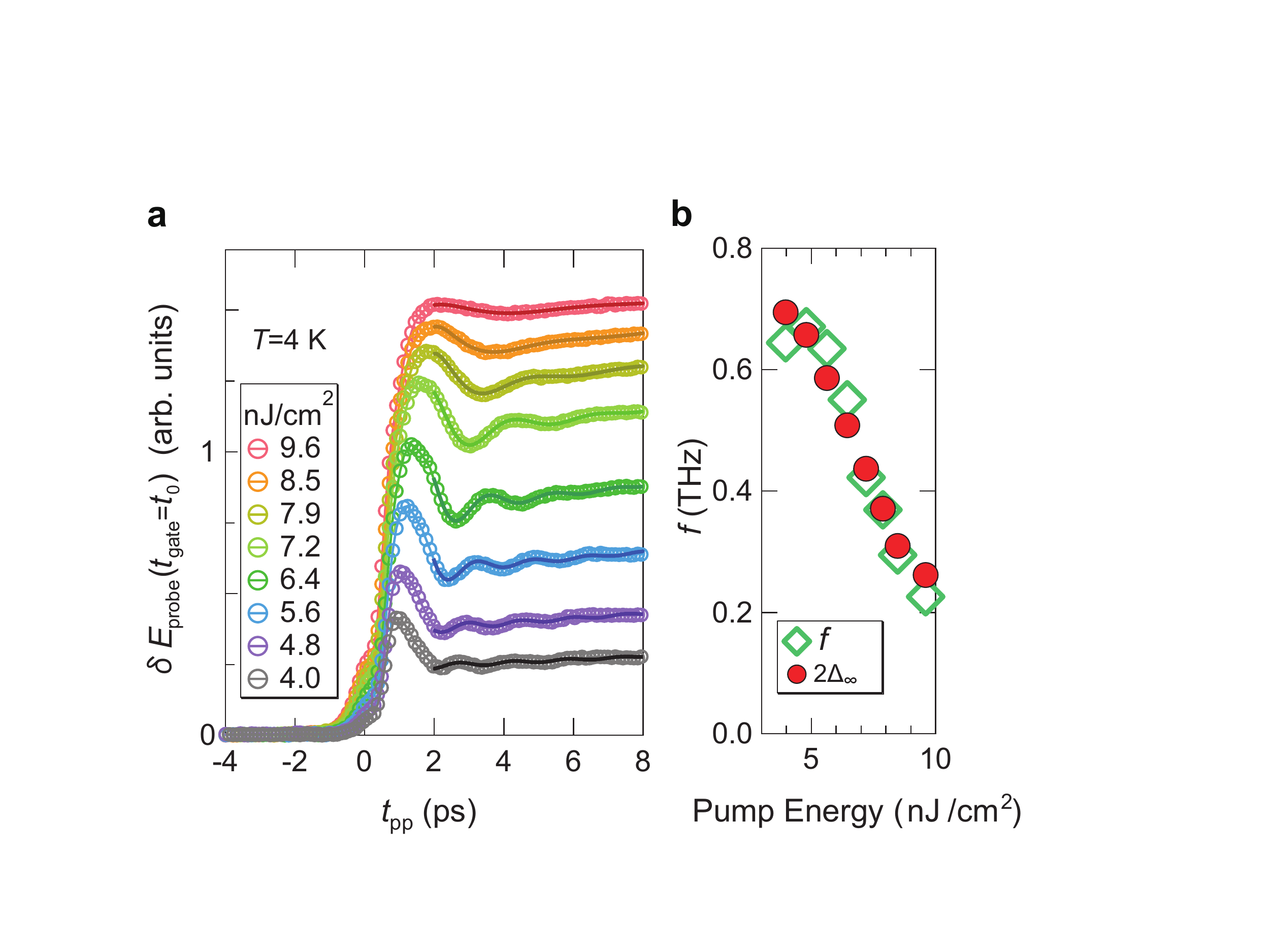}
\caption{Higgs-mode oscillation after the nonadiabatic excitation of quasiparticles in NbN induced by a monocycle THz pump. ({\it a}) The temporal evolution of the change of the probe electric field, $\delta E_{\rm probe}$, as a function of the pump-probe delay time $t_{\rm pp}$ at various pump intensities. The solid curves represent the results fitted by a damped oscillation with a power-law decay. ({\it b}) The oscillation frequency $f$ obtained from the fits and the asymptotic gap energy $2\Delta_{\infty}$ as a function of the pump intensity. Figures are taken from Reference~\cite{Matsunaga2013}.}
\label{Higgs oscillation}
\end{center}
\end{figure}

Despite intensive theoretical studies on the Higgs mode in superconductors, the observation of the Higgs mode has long been elusive except for the case of 2H-NbSe$_2$, until when the ultrafast THz-pump and THz-probe spectroscopy was performed in a conventional $s$-wave superconductor Nb$_{1-x}$Ti$_x$N \cite{Matsunaga2013}.
In this experiment, quasiparticles are instantaneously injected at the superconducting-gap edge 
to quench the superconducting order parameter nonadiabatically. 
A strong single-cycle THz pump pulse with the center frequency around 1 THz ($\sim$ 4 meV) was generated from a LiNbO$_3$ crystal by using the tilted-pulse-front method \cite{Hebling2008, Watanabe2011, Shimano2012}, in order to excite high density quasiparticles just above the gap of Nb$_{1-x}$Ti$_x$N thin films with $2\Delta=$ 0.72-1.3 THz ($\sim$ 3.0-5.4 meV) at 4 K. The details of the THz pump-THz probe scheme were given in Ref. \cite{Matsunaga2012}. The subsequent dynamics of the superconducting order parameter was probed by a weaker probe THz pulse in transmission geometry. {\bf Figure} \ref{Higgs oscillation} shows pump-probe delay dependence of the transmitted probe THz electric field at a fixed point of the probe pulse waveform which reflects the order parameter dynamics. A clear oscillation is identified after the pump with the oscillation frequency given by the asymptotic value of the superconducting gap $2\Delta_\infty$ caused by the quasiparticle injection. The decay of the oscillation is well fitted by a power-law decay predicted for the Higgs-mode oscillation. With increasing the pump intensity, the oscillation period shows a softening, as the asymptotic value of $2\Delta_\infty$ becomes smaller with higher quasiparticle densities. 
By recording the probe THz waveform at each pump-probe delay, one can extract the dynamics of the complex optical conductivity $\sigma(\omega)=\sigma_1(\omega)+i\sigma_2(\omega)$ (not shown).
The spectral-weight oscillation with frequency $2\Delta_\infty$ was clearly identified both in the real and imaginary parts \cite{Matsunaga2017b},
the latter of which reflects the oscillation of the superfluid density (i.e., the amplitude of the superconducting order parameter). 

The intense single-cycle THz pump pulse was crucial to excite the quasiparticles instantaneously at the gap edge which acts as a nonadiabatic quench of the order parameter, thereby inducing a free oscillation of the Higgs mode. On the contrary, when a near-infrared optical pulse was used as a pump, a relatively slow rise time ($\sim$ 20 ps) was observed for the suppression of the order parameter, though dependent on the excitation fluence \cite{Matsunaga2012, Beck2011}. In this case, the photoexcited hot electrons (holes) emit a large amount of high frequency phonons ($\omega>2\Delta$) (HFP) which  subsequently induces the Cooper pair-breaking until quasiparticles and HFP reach the quasiequilibrium.
This process was well described by the Rothwarf-Taylor phonon-bottleneck model \cite{Beck2011}, and usually takes a time longer than $(2\Delta)^{-1}$, which breaks the nonadiabatic excitation condition needed for the quench experiment. It should be noted here that even if one uses the near-infrared or visible optical excitation, the instantaneous quasiparticle excitation can occur if the stimulated Raman process is efficient \cite{Mansart2013}.

The decay of the Higgs mode is also an important issue. Since the Higgs-mode energy $2\Delta$ coincides with the onset of quasiparticle continuum, the Higgs mode inevitably decays into the quasiparticle continuum even without any collisions \cite{Volkov1974, Kulik1981}. However, this does not mean that the Higgs mode is an overdamped mode. In fact, the decay due to the collisionless energy transfer is expected to be proportional to $t^{-1/2}$ in the BCS superconductor, ensuring a much longer lifetime compared with the oscillation period. The energy dispersion of the Higgs mode for finite wavenumber $q$ has been calculated as 
\begin{align}
\omega_H
&\simeq
2\Delta+\frac{v_F^2}{12\Delta}q^2-i\frac{\pi^2v_F}{24}q
\label{dispersion}
\end{align}
where $v_F$ is the Fermi velocity \cite{Littlewood1982}. From Equation~\ref{dispersion} one can estimate that as the wavenumber $q$ increases and the mode energy enters into the quasiparticle continuum, the lifetime of Higgs mode is shortened and turns into an overdamped mode. 
For the case of NbN,
$\Delta\sim$ 0.6 THz and $v_F \sim 2 \times 10^6$ m/s \cite{Chockalingam2008}
and the imaginary part exceeds the real part in Equation~\ref{dispersion} at $q>\Delta/v_F \sim$ 0.3 ($\mu$m)$^{-1}$. 
In the single-cycle THz-pump experiment in a thin film of NbN \cite{Matsunaga2013}, the value of in-plane wavenumber $q$ estimated from the spot size of the pump ($\sim$ 1 mm) is at most $\sim$ 1 (mm)$^{-1}$, which is far smaller than $\Delta/v_F$. In such a small $q$ limit, the Higgs mode is considered as a well-defined collective mode \cite{Littlewood1982}.

\subsubsection{Multicycle THz driving with subgap frequency and third harmonic generation}

\begin{figure}
\begin{center}
\includegraphics[width=12cm]{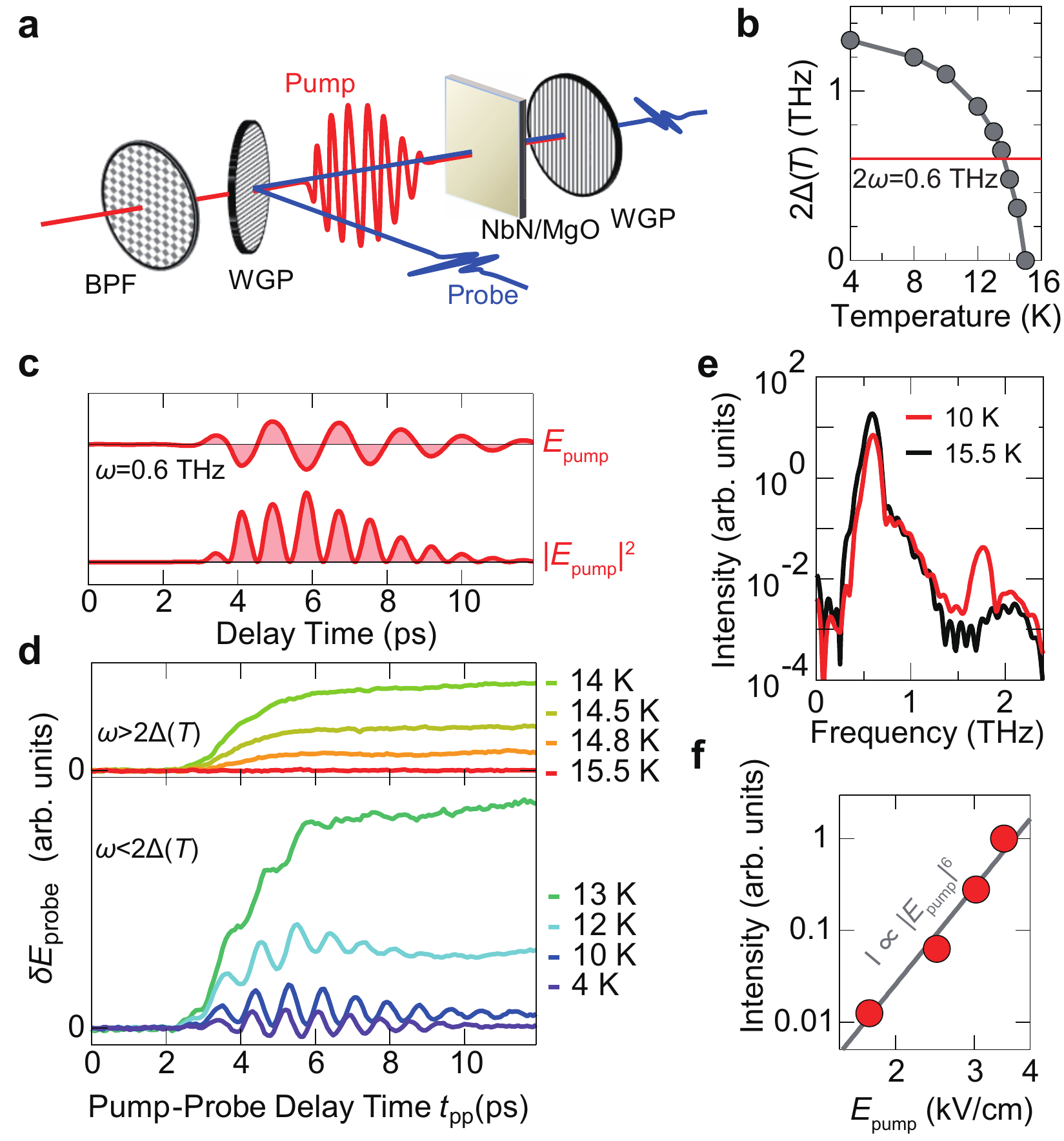}
\caption{Forced Higgs oscillation and third harmonic generation from a superconducting NbN film under the multicycle THz pump. ({\it a}) Schematic setup for the multicycle THz pump and THz probe spectroscopy. BPF: a bandpass filter, WGP: a wire grid polarizer. ({\it b}) Temperature dependence of the superconducting gap energy. Horizontal line indicates the center frequency of the pump pulse, $\omega=0.6$ THz. ({\it c}) Waveform of the pump THz electric field $E_{\rm pump}$ and  the squared one, $|E_{\rm pump}|^2$. ({\it d}) The change in the transmitted probe THz electric field $\delta E_{\rm probe}$ as a function of the pump-probe delay time $t_{pp}$ at the tempareture range  $\omega>2\Delta (T)$ (top panel) and $\omega<2\Delta (T)$ (bottom panel). Increase of $\delta E_{\rm probe}$ corresponds to the reduction of the order parameter. ({\it e}) Power spectra of the transmitted pump THz pulse above and below $T_c=15$ K. ({\it f}) THG intensity as a function of the pump THz field strength. Figures are taken from Reference~\cite{Matsunaga2014}.}
\label{Multicycle}
\end{center}
\end{figure}

To excite the Higgs mode in an on-resonance condition,
one can use a narrowband multicycle THz pump pulse with the photon energy tuned below the superconducting gap $2\Delta$.
With this, the amplitude of the superconducting order parameter was shown to oscillate 
with twice the frequency of the incident pump-pulse frequency $\omega$ during the pulse irradiation
({\bf Figure}~\ref{Multicycle}{\textit{\textbf a}}-{\textit{\textbf d}}) \cite{Matsunaga2014}. The observed $2\omega$ modulation of the order parameter was attributed to the forced oscillation of the Higgs mode caused by electromagnetic fields. In fact, the time-dependent Ginzburg-Landau theory shows that there is a coupling term between the Higgs ($H$) and field ($\bm A$) as expressed by $\bm A^2 H$ (Section~\ref{phenomenological view}), 
which results in the $2\omega$-order parameter modulation.
The nonlinear electromagnetic response of superconductors has been studied
in the frequency range far below $2\Delta$ and near $T_c$ where the Ginzburg-Landau theory is applicable \cite{Abraham1966, Gorkov1968, Gorkov1969, Amato1976, Entin-Wohlman1978}.
In the theory developed by Gor'kov and Eliashberg (GE), a closed set of equations for the order parameter $\Delta$ and the field $\bm A$ was derived, which predicts the $2\omega$ modulation of $\Delta$ and the THG from the induced supercurrent $\bm j \propto \bm A^2 \Delta$ \cite{Gorkov1968}. The GE theory was experimentally demonstrated through the observation of THG of a microwave field in a paramagnetically doped superconductor \cite{Amato1976}.

\begin{figure}
\begin{center}
\includegraphics[width=12cm]{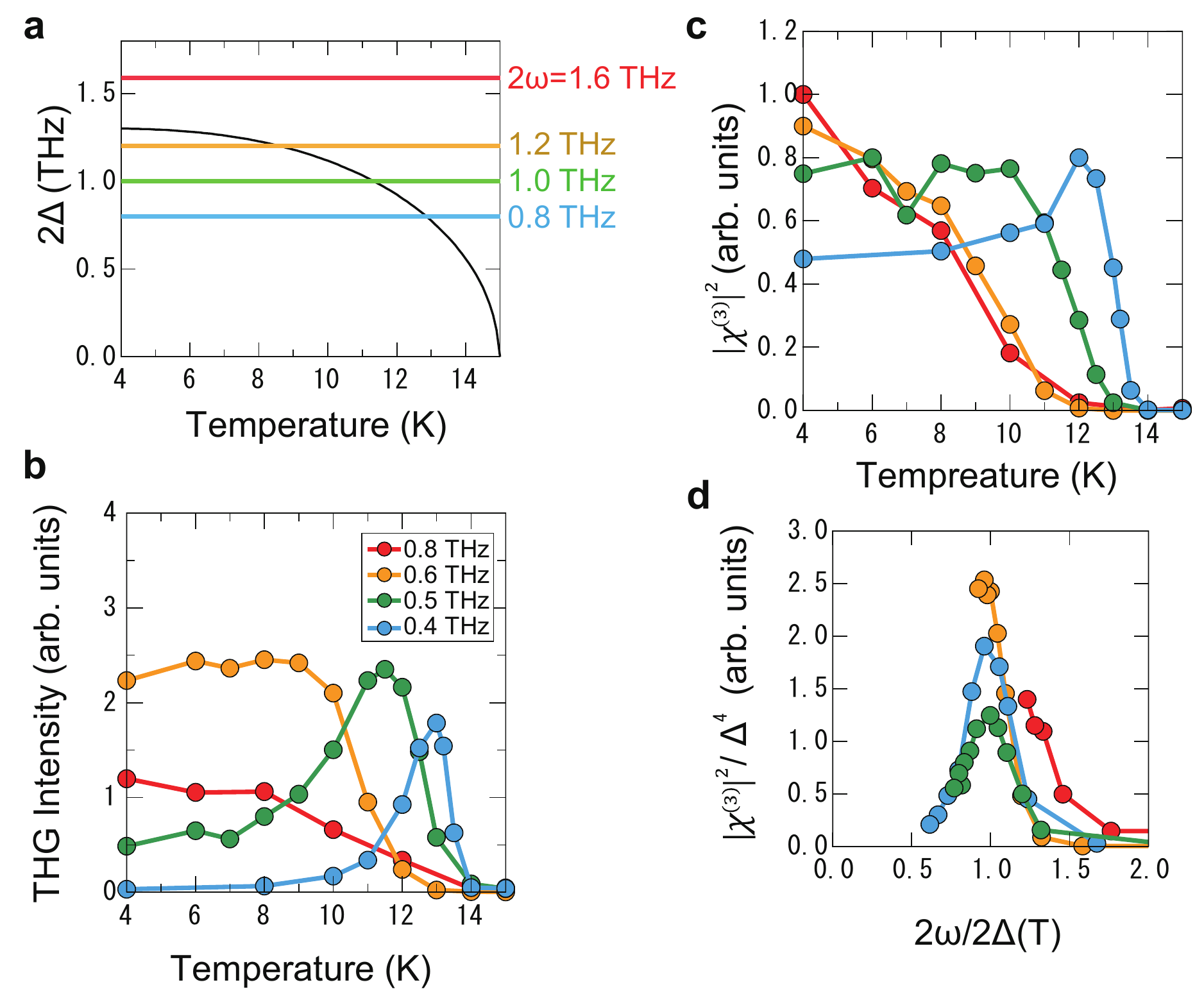}
\caption{Temperature dependence of THG signal from NbN. ({\it a}) Temperature dependence of the supercondcuting gap $2\Delta(T)$. Twice the THz pump frequencies $2\omega$ are shown by horizontal lines. ({\it b}) Measured THG intensities at each frequencies as a function of temperature. ({\it c}) THG intensities normalized by the temperature dependent internal field inside the superconducting film and the transmittance of the THG signal, which corresponds to the susceptibility of THG, $|\chi^{(3)}|^2$. ({\it d}) Further normalized value of $|\chi^{(3)}|^2$ by the temperature dependent gap $\Delta(T)^4$ as a function of normalized frequency $2\omega/2\Delta(T)$. Relative intensities between different frequencies are arbitrary. Data are taken from Reference~\cite{Matsunaga2017b}.}
\label{THG resonance}
\end{center}
\end{figure}

The THz pump experiments have shed a new light on the investigation of the nonlinear electromagnetic response of superconductors at the gap frequency region. Subsequent to the observation of the $2\omega$ modulation of the order parameter, namely the forced oscillation of the Higgs modes, the THG was observed under the irradiation of a subgap multicycle THz pump pulse in NbN thin films as shown in {\bf Figure}~\ref{Multicycle}{\textit {\textbf e}}, {\textit{\textbf f}}. Importantly, the THG was shown to be resonantly enhanced when a condition $2\omega=2\Delta$ is satisfied \cite{Matsunaga2014}. The temperature dependence of THG intensity for various frequencies in a NbN film is shown in {\bf Figure}~\ref{THG resonance}: the temperature dependence of the gap $2\Delta$ and the incident frequencies ({\bf Figure}~\ref{THG resonance}{\textit{\textbf a}}), the raw data of THG intensities ({\bf Figure}~\ref{THG resonance}{\textit{\textbf b}}), those normalized by the internal field inside the superconducting film which is proportional to the third-order susceptibility $|\chi^{(3)}|^2$ ({\bf Figure}~\ref{THG resonance}{\textit{\textbf c}}), and the THG intensity further normalized by the proportional factor of the THG intensity $\Delta(T)^4$ versus the frequency normalized to the gap, $2\omega/2\Delta(T)$ ({\bf Figure}~\ref{THG resonance}{\textit{\textbf d}}). These results clearly show the two-photon resonances of the Higgs mode with the field. 

As previously described in Section \ref{impurity}, however, it was theoretically pointed out that there is a contribution also from the quasiparticle excitation (or called charge density fluctuations) in THG, which also exhibits the $2\omega=2\Delta$ resonance \cite{Cea2016}. Within the BCS mean-field approximation in the clean limit, the quasiparticle term was shown to exceed the Higgs mode contribution in THG. At the same time, it was predicted that the THG from the quasiparticle term should exhibit polarization dependence with respect to the crystal axis while the Higgs term is totally isotropic in a square lattice \cite{Cea2016}. In the experiments in NbN, the THG intensity was shown to be independent from the incident polarization angle with respect to the crystal orientation and notably the emitted THG does not have orthogonal components with respect to the incident polarization of the fundamental ($\omega$) wave \cite{Matsunaga2017}. Such a totally isotropic nature of THG cannot be well accounted for by the CDF term even if one considers the crystal symmetry of NbN \cite{Cea2018} and indicates the Higgs mode as the origin of THG. Recently, the importance of $\bm p\cdot \bm A$ term on the coupling between the Higgs mode and the gauge field was elucidated. In particular, the $\bm p\cdot \bm A$ term contribution is largely enhanced when one consider the phonon retardation effect beyond the BCS approximation \cite{Tsuji2016} and more prominently when one considers the nonmagnetic impurity scattering effect \cite{Jujo2018, Murotani2019, Silaev2019}, dominating the quasiparticle term by an order of magnitude as described in Section \ref{impurity}. This paramagnetic term was also shown to dominate the diamagnetic term, 
although its ratio depends on the ratio between the coherence length and the mean free path. The NbN superconductor is in the dirty regime as manifested by a clear superconducting gap structure observed in the conductivity spectrum, and the paramagnetic term should significantly contribute to THG.

\subsection{Higgs mode in high-$T_c$ cuprates}
\label{cuprate}

\begin{figure}
\begin{center}
\includegraphics[width=10cm]{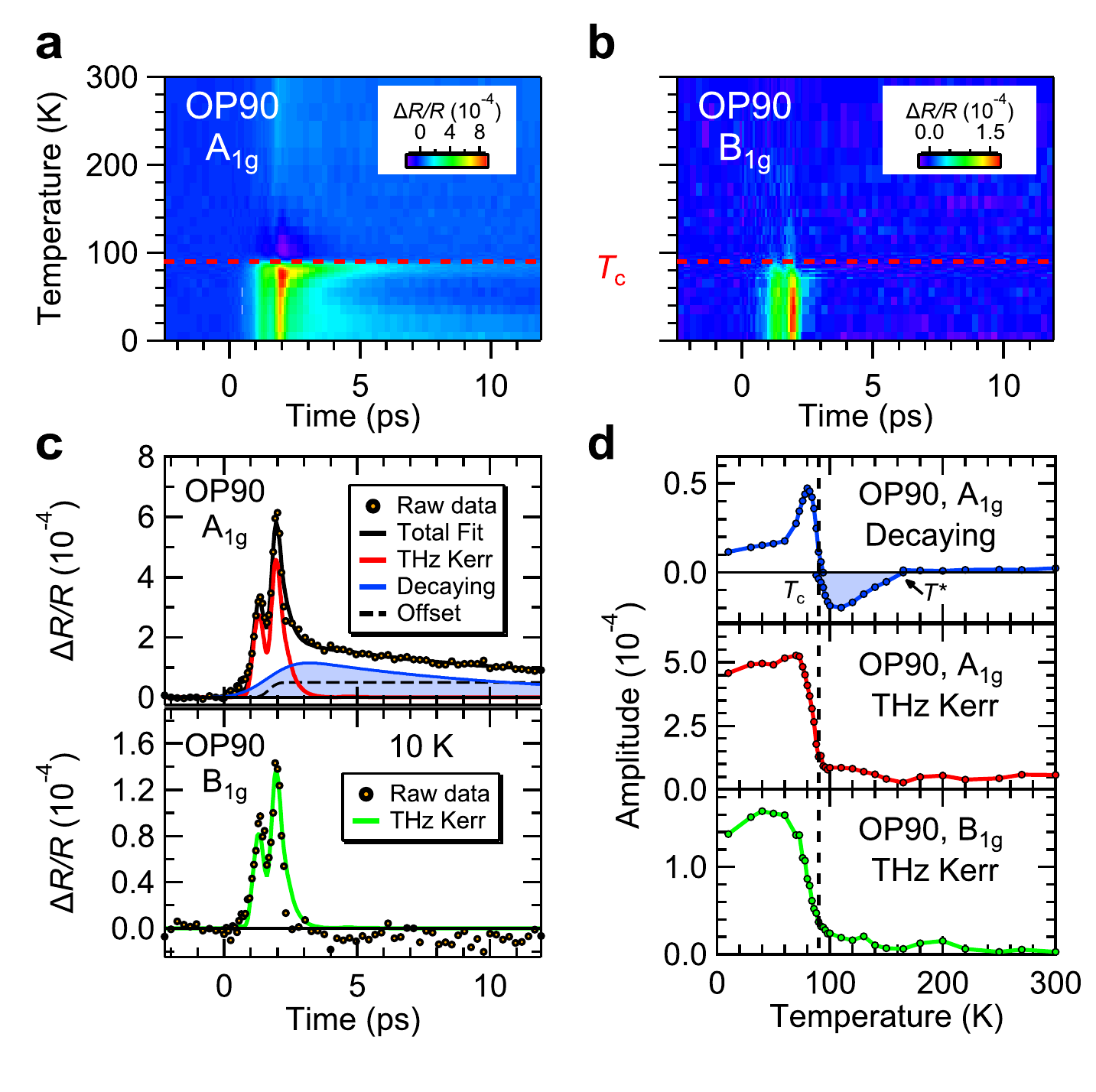}
\caption{Forced oscillation of Higgs mode in an optimally doped (OP90) Bi$_2$Sr$_2$CaCu$_2$O$_{8+x}$ measured by THz pump and optical reflection probe experiment. Temperature dependences of reflectivity change $\Delta R/R$ as a function of pump-probe delay time in ({\it a}) A$_{\rm 1g}$ and ({\it b}) B$_{\rm 1g}$ components. Red dashed lines indicate $T_c$. ({\it c}) The A$_{\rm 1g}$ and B$_{\rm 1g}$ components against the pump-probe delay time at 10 K. Fitting curves are also shown by solid lines. ({\it d}) Temperature dependences of the A$_{\rm 1g}$ decaying component (blue) (incoherent quasiparticle excitation), the A$_{\rm 1g}$ oscillatory component (red) (forced Higgs oscillation), and the B$_{\rm 1g}$ oscillatory component (green) (likely the charge density fluctuation). Figures are taken from Reference~\cite{Katsumi2018}.}
\label{Bi2212}
\end{center}
\end{figure}

Higgs modes can exist not only in $s$-wave superconductors but also in unconventional superconductors. For example, in $d$-wave superconductors such as high-$T_c$ cuprates one can expect various types of collective modes, including not only the totally isotropic oscillation of the gap function in relative momentum space (A$_{\rm 1g}$ mode) but also those that oscillate anisotropically (e.g., A$_{\rm 2g}$, B$_{\rm 1g}$, B$_{\rm 2g}$ modes) \cite{Barlas2013}. These are reminiscent of Bardasis-Schrieffer mode for the case of $s$-wave superconductors \cite{Bardasis1961}. Possible collective modes have been classified based on the point group symmetry of the lattice \cite{Barlas2013}, and the relaxation behavior of the A$_{\rm 1g}$ mode induced by a quench has been discussed \cite{Peronaci2015}. A theoretical proposal has been made to observe the Higgs mode in a $d$-wave superconductor through a time- and angle-resolved photoemission spectroscopy \cite{Nosarzewski2017}.

The observation of the Higgs mode in $d$-wave superconductors was recently made in THz-pump and optical-probe experiments in high-$T_c$ cuprates, Bi$_2$Sr$_2$CaCu$_2$O$_{8+x}$ (Bi2212) \cite{Katsumi2018}. An oscillatory signal of the optical reflectivity that follows the squared THz electric field was observed, which is markedly enhanced below $T_c$ as shown in {\bf Figure}~\ref{Bi2212}. This signal was interpreted as the THz-pump induced optical Kerr effect, namely the third-order nonlinear effect induced by the intense THz pump pulse. In the Bi2212 system, both A$_{\rm 1g}$ and B$_{\rm 1g}$ symmetry components with respect to
the polarization-angle dependence (which should be distinguished from the relative-momentum symmetry classification mentioned above)
were observed in the THz-Kerr signal.
The doping dependence shows that the A$_{\rm 1g}$ component is dominant in all the measured samples from under to near-optimal region. From the comparison with the BCS calculation of the nonlinear susceptibility, the A$_{\rm 1g}$ component was
assigned to be the $d$-wave Higgs-mode contribution.

The observed oscillatory signal corresponds to the $2\omega$ modulation of the order parameter as observed in an $s$-wave NbN superconductor under the subgap multicycle THz pump irradiation. Accordingly, like in the case of $s$-wave systems, one can expect the THG signal from high-$T_c$ cuprates. Recently, THG was indeed observed ubiquitously in La$_{1.84}$Sr$_{0.16}$CuO$_4$, DyBa$_2$Cu$_3$O$_{7-x}$, YBa$_2$Cu$_3$O$_{7-x}$, and overdoped Bi$_2$Sr$_2$CaCu$_2$O$_{8+x}$ \cite{Chu2019} by using 0.7 THz coherent and intense light source at the TELBE beamline at HZDR. 

\subsection{Higgs mode in the presence of supercurrents}
\label{supercurrent}

So far, we have seen the Higgs mode excitation by the nonadiabatic quench with instantaneous quasiparticle injection 
or through the quadratic coupling with the electromagnetic wave. Recently, it has been theoretically shown that
when there is a condensate flow (i.e., supercurrent), the Higgs mode linearly interacts with electromagnetic waves with the electric field polarized along the direction of the supercurrent flow \cite{Moor2017}. 
Namely, the Higgs mode should be observed in the optical conductivity spectrum at the superconducting gap $\omega=2\Delta$ 
under the supercurrent injection. This effect is explained by the momentum term in the action,
$S\propto\int \bm Q^2(t)|\Delta(t)|^2 dt dr$,
where $\bm Q(t)=\bm Q_0+\bm Q_\Omega(t)$ is the gauge-invariant momentum of the condensate,
$\bm Q_0$ is the dc supercurrent part, and $\bm Q_\Omega(t)={\rm Re}[\bm Q_\Omega \exp(i\Omega t)]$ is the time-dependent part driven by the ac probe field, respectively. 
$\Delta(t)=\Delta_0+\delta\Delta(t)$ is the time-dependent superconducting order parameter. 
The action $S$ includes the integral of $\delta\Delta_{2\Omega} \bm Q^2_{-\Omega}$
and $\delta\Delta_\Omega \bm Q_0 \bm Q_{-\Omega}$ 
where $\delta\Delta_{\Omega(2\Omega)}$ is the Fourier component of the oscillating order parameter. 
The first term corresponds to the quadratic coupling of the Higgs mode to the gauge field as described Section~\ref{nonlinear coupling}.
The second term indicates that the Higgs mode linearly couples with the gauge field under the supercurrent with the condensate momentum $\bm Q_0$ parallel to the probe electric field.

\begin{figure}
\begin{center}
\includegraphics[width=12cm]{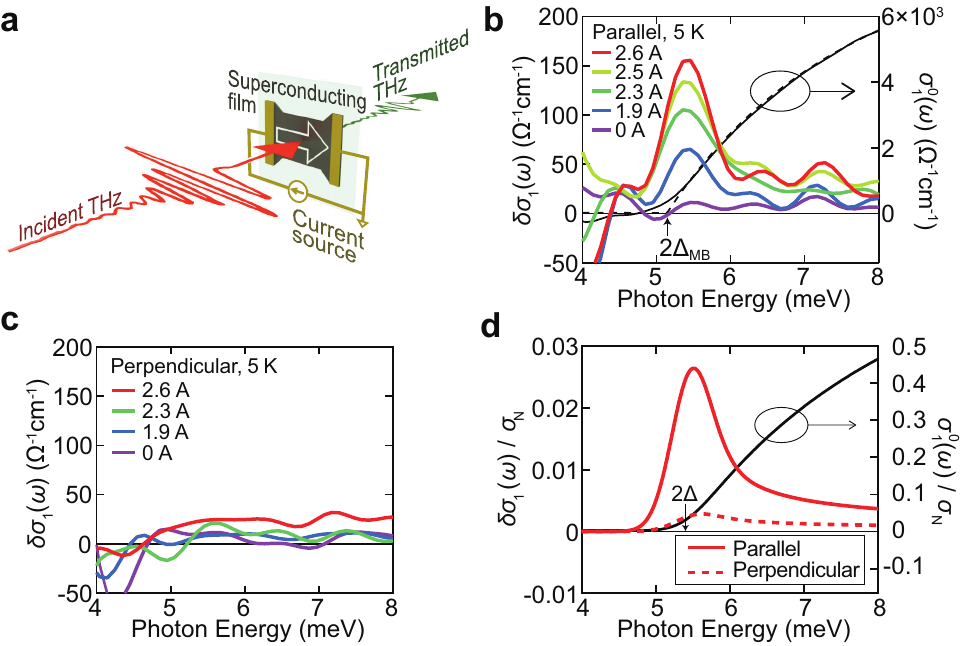}
\caption{({\it a}) Schematic view of the terahertz transmittance experiments under supercurrent injection. The change of the optical conductivity induced by supercurrent injection taken with the THz probe electric field ({\it b}) parallel or ({\it c}) perpendicular to the current direction. The optical conductivity spectra measured without the current is also plotted in ({\it b}). The dashed line represents a fit by the Mattis-Bardeen model \cite{Mattis1958, Zimmermann1991}, and the superconducting gap estimated from the fit is indicated by the vertical arrow. ({\it d}) Theoretically expected changes of optical conductivity induced by the supercurrent injection with respect to the normal state conductivity $\sigma_N$. Figures are taken from Reference~\cite{Nakamura2018}.}
\label{MovingCondensate}
\end{center}
\end{figure}

Recently, in accordance with the theoretical prediction, the Higgs mode was observed in the optical conductivity in an $s$-wave superconductor NbN thin film under the supercurrent injection by using THz time-domain spectroscopy \cite{Nakamura2018}.
{\bf Figure} \ref{MovingCondensate}{\textit{\textbf a}} shows the schematic experimental setup. 
{\bf Figures} \ref{MovingCondensate}{\textit{\textbf b}} and {\textit{\textbf c}} show the change of the real part of the optical conductivity induced by the supercurrent injection with the probe polarization ({\textit{\textbf b}}) parallel and ({\textit{\textbf c}}) orthogonal to the current direction. A clear peak is observed at the gap edge $\omega=2\Delta$ in the parallel geometry, showing a good agreement with the spectra calculated on the basis of the theory developed by Moor et al. \cite{Moor2017} as represented in {\bf Figure} \ref{MovingCondensate}{\textit{\textbf d}}.
This result indicates that, without resorting to the sophisticated nonlinear THz spectroscopy, the Higgs mode can be observed in the linear response function when the condensate has a finite momentum, expanding the feasibility for the detection of the Higgs mode in a wider range of materials. 

The visibility of the Higgs mode in the optical conductivity has also been addressed in a two-dimensional system near a quantum critical point \cite{Podolsky2011, Gazit2013} (see also \cite{Sachdev1999, Zwerger2004}).
Experimentally, an extra spectral weight in the optical conductivity below the superconducting gap was observed in a strongly disordered superconducting film of NbN,
which was interpreted in terms of the Higgs mode with its mass pushed below the pair breaking gap due to the strong disorder \cite{Sherman2015}. However, the assignment of the extra spectral weight below the gap remains under debates. 
It can also be described by the Nambu-Goldstone mode which acquires the electric dipole in strongly disordered superconductors \cite{Cea2014b, Cea2015, Pracht2017, Seibold2017}. The extra optical conductivity was also accounted for by the disorder induced broadening of the quasiparticle density of states below the gap \cite{Cheng2016}.

\section{FUTURE PERSPECTIVE}

Having seen the Higgs mode in $s$-wave and $d$-wave superconductors, it is fascinating to extend the observation of the Higgs mode to a variety of conventional/unconventional superconductors. 

The study of the Higgs mode in multiband superconductors would be important, as it would give deeper insights into the interband couplings. For instance, the application of the nonlinear THz spectroscopy to iron-based superconductors is highly intriguing, as it may provide information on the interband interactions and pairing symmetry \cite{Akbari2013, Maiti2015, Muller2018}. The Leggett mode, namely the collective mode associated with the relative phase of two condensates \cite{Leggett1966} is also expected to be present in multiband superconductors. 
The Leggett mode has been observed in a multiband superconductor MgB$_2$ by Raman spectroscopy \cite{Blumberg2007}. 
The nonlinear coupling between the Leggett mode and THz light has also been discussed \cite{Murotani2019, Akbari2013, Krull2016, Cea2016b, Murotani2017}.
Recently, a THz-pump THz probe study in MgB$_2$ has been reported, and the results were interpreted in terms of the Leggett mode \cite{Giorgianni2019}. However, the dominant contribution of the Higgs mode over the Leggett mode in the nonlinear THz responses was theoretically pointed out in the case of dirty-limit superconductors \cite{Murotani2019}. 
Therefore, further experimental verification of the Higgs and Leggett modes is left as a future problem. 

The ultimate fate of the behavior of the Higgs mode in a strongly correlated regime and, what is more,
in a BCS-BEC crossover regime is an intriguing problem. 
In particular, the decay profile of the Higgs mode has been predicted to change from the BCS to BEC regime \cite{Gurarie2009, Tsuji2013, Scott2012, Yuzbashyan2015, Tokimoto2019}. Experimentally, such a study has been realized in a cold-atom system, showing a broadening of the Higgs mode in the BEC regime \cite{Behrle2018}. Further investigation on the Higgs mode in the BCS-BEC crossover regime will be a future challenge in solid-state systems. 

The Higgs mode in spin-triplet $p$-wave superconductors (whose candidate materials include Sr$_2$RuO$_4$) is another issue to be explored in the future. Like in the case of superfluid $^3$He \cite{Volovik2014, He3book, Wolfle1977}, multiple Higgs modes are expected to appear due to the spontaneous symmetry breaking in the spin channel, and its comparison with high-energy physics would be interesting. 

As yet another new paradigm, the observation of the Higgs mode would pave a new pathway for the study of nonequilibrium phenomena, in particular for the photoinduced superconductivity \cite{Fausti2011, Kaiser2014, Hu2014, Mitrano2016}. Being a fingerprint of the order parameter with a picosecond time resolution, the observation of the Higgs mode in photoinduced states should offer a direct evidence for nonequilibrium superconductivity. In strongly correlated electron systems exemplified by the unconventional superconductors, elucidation of the interplay between competing and/or coexisting orders with superconductivity is an important issue to understand the emergent phases. The time-domain study of collective modes associated with those orders is expected to provide new insights for their interplay. 

\section{DISCLOSURE STATEMENT}
The authors are not aware of any affiliations, memberships, funding, or financial holdings
that might be perceived as affecting the objectivity of this review.

\section*{ACKNOWLEDGMENTS}
We wish to acknowledge valuable discussions with Hideo Aoki, Yann Gallais, Dirk Manske, Stefan Kaiser, and Seiji Miyashita.
We also acknowledge coworkers, Yuta Murotani, Keisuke Tomita, Kota Katsumi, Sachiko Nakamura, Naotaka Yoshikawa, and in particular Ryusuke Matsunaga for 
fruitful discussions and their support for preparing the manuscript.

%

\end{document}